\documentclass[prl,aps,amsfonts,amsmath,superscriptaddress,%
  twocolumn,showpacs,floatfix]{revtex4-1}
\usepackage{amsfonts}
\usepackage{amsmath}
\usepackage{bm}
\usepackage{epsfig}
\usepackage{mathrsfs}

\usepackage[usenames]{color}
\usepackage{ulem}

\begin{document}

 \title {\bf Dimensional regularization applied to nuclear matter with a zero-range interaction}
\author{Kassem Moghrabi}
\address{Institut de Physique Nucl\'eaire,
 Universit\'e Paris-Sud, IN2P3-CNRS, F-91406 Orsay Cedex, France}
\author{Marcella Grasso}
\address{Institut de Physique Nucl\'eaire,
 Universit\'e Paris-Sud, IN2P3-CNRS, F-91406 Orsay Cedex, France}


\begin{abstract}
We apply the dimensional regularization procedure to 
treat an ultraviolet divergence occurring in the framework of the nuclear many-body 
problem. We consider
the second--order correction (beyond the mean--field approximation) to the 
equation of state of nuclear matter with a zero--range 
effective interaction. The unphysical ultraviolet divergence that is generated 
at second order by the zero range of the interaction is removed 
by the  regularization technique and the regularized equation of state (mean--field 
+ second--order contributions) is adjusted to a reference equation of state. 
The main  practical advantage of this procedure, with respect to a 
cutoff regularization, is to provide a unique set of parameters for the 
adjusted effective interaction. This occurs because the regularized second--order 
correction does not contain any cutoff dependence. The encouraging results found in this work indicate that such an elegant 
technique to generate regularized 
effective interactions is likely to be applied in future to finite nuclei in 
the framework of beyond mean--field models.

\end{abstract}

\vskip 0.5cm  \maketitle


\section{Introduction}
The mean--field framework in many-body techniques corresponds to the first--order term of the perturbative solution of the Dyson equation. In several  beyond mean--field models higher--order contributions of this perturbative expansion are also taken into account to enrich the theoretical description.
Perturbation theories are employed in many domains of physics and  regularization and renormalization techniques are 
 adopted in  cases where the inclusion in perturbative expansions of higher--order terms with respect to the leading contribution generates divergences. This type of 
divergences are well known for instance in particle physics in the context of quantum field theories \cite{peskin}.
In
several frameworks of the many--body physics, such as nuclear  
and atomic physics, other types of divergences occur in 
mean--field--based models  
(for example, the Bogoliubov--de Gennes or the Hartree--Fock--Bogoliubov theories) if a 
zero--range interaction is employed in the pairing channel to treat a superfluid many--fermion 
system \cite{bruun,bulgac,grasso}. Apart from this specific case related to the pairing interaction, in the perturbative treatment of the many--body problem 
an ultraviolet divergence 
always appears, if a zero--range interparticle interaction is used, 
when
higher--order terms are included beyond the mean-field approximation. 
This unphysical divergence has been analyzed in previous works \cite{mog2010,mog2012}. 

We apply here a technique that is currently adopted in the context of 
quantum field theories, the so--called dimensional regularization, for the second--order correction beyond the mean--field approximation within a specific case of the nuclear many--body problem. In Ref. 
\cite{mog2010}, we have considered the equation of state (EoS) of symmetric nuclear matter with a contact interaction $g(\rho) \; \delta(\vec{r}_1-\vec{r}_2)$ where the coupling constant $g(\rho)$ depends on the density $\rho$ and contains three parameters, $t_0$, $t_3$ and $\alpha$: $g(\rho)=t_0+\frac{t_3}{6} \,\rho^{\alpha}$. This corresponds to the $t_0-t_3$ model of the nuclear effective Skyrme interaction \cite{skyrme}. In Ref. \cite{mog2010} we have analyzed in this specific model the nature of the divergence related to the second--order correction of the EoS. The divergent term has been calculated analytically. By deriving its asymptotic expansion, it is possible to show that it diverges linearly with the momentum $\Lambda$ which is introduced as a cutoff regulator. A cutoff regularization has been applied to absorb the divergence by means of a fit of parameters in the corrected 
mean--field + second--order EoS. 
The same cutoff technique has been applied in Ref. \cite{mog2012} to an enriched model where the velocity--dependent terms have also been included in the Skyrme interaction. This modifies the character of the divergent term that is in this case proportional to $\Lambda^5$. Not only symmetric, but also asymmetric and 
pure neutron matter have been considered in Ref. \cite{mog2012} and a fit of parameters has been performed for each value of the cutoff as in Ref. \cite{mog2010} .

In this work, we remove the divergent term of the second--order correction 
(linear in the cutoff for the $t_0-t_3$ model and proportional to $\Lambda^5$ for the full Skyrme interaction)
 with the dimensional regularization technique that is applied here to the second--order EoS of nuclear matter.  
We mention that the dimensional regularization technique has been already applied in the past to dilute Fermi systems in the context of effective field theories (see, for instance, Refs. \cite{hammer,kaiser}).
This regularization technique is an elegant procedure which has the advantage of preserving symmetry laws and of including high--energy effects which are sharply discarded with a cutoff regularization. It has been introduced in the framework of the electroweak theory \cite{hoo1,hoo2,boll} and consists in replacing the dimension of the divergent integrals with a continuous variable $d$. The main idea is that, if an integral diverges in a given integer dimension, the result may be finite by replacing the integer dimension with a non integer $d$. One then performs  a kind of analytic continuation in the dimension to return to the initial integer value \cite{wilson}. The dimensional regularization eliminates power--law divergences, isolates logarithmic divergences and regularizes infrared divergences \cite{leibb}. 
Together with the continuous variable $d$, a regulator $\epsilon$ (which is dimensionless) is introduced so that, when $\epsilon \rightarrow$  0, the dimension of the integral comes back to the initial integer value. In addition, an auxiliary scale $\mu$ is included to maintain the correct dimensions of the physical quantities which are calculated. After regularization, a renormalization is applied by a minimal subtraction procedure to remove the regulator $\epsilon$ which appears as a $1/\epsilon$ pole if the divergence is logarithmic. All the physical observables are independent of the auxiliary scale $\mu$ due to the renormalization group equation 
$\mu dS/d\mu =0$, where $S$ is a generic observable.

After having derived the regularized second--order correction for the nuclear EoS (wich is now finite and independent of the cutoff) 
we follow the same procedure as in our previous works \cite{mog2010,mog2012} and we adjust the corrected EoS to reproduce a reasonable EoS chosen as a reference. This time, we generate a unique set of parameters whereas in the previous works a set of parameters was found for each value of the cutoff. 

The article is organized as follows. In Sec. II the regularization procedure is presented 
first 
for the $t_0-t_3$ model and then for the complete Skyrme interaction. 
In the first case more details are provided for the analytical derivation. For simplicity, only the final expressions are reported for the 
second case. 
In Sec. III the results for the fit of the parameters are discussed and in Sec. IV some conclusions and prospectives are summarized.

\section{Dimensional regularization tecnhique for the second-order EoS}

\subsection{Regularized EoS for symmetric matter: $t_0-t_3$ model}

Let us consider the second--order correction beyond the mean--field EoS in the $t_0-t_3$ model for symmetric nuclear matter \cite{mog2010}. We write its generalized expression where we introduce a continuous dimension $d$ in the integral and the auxiliary scale $\mu$. In a box of volume $\Omega$ one has, 
\begin{widetext}
\begin{eqnarray}\label{eqn1}
\frac{\Delta E(\rho)}{A}&=&\frac{6}{\mu^{3(d-3)}}\frac{\Omega^{d-3}}{(2\pi)^{3d}}\left(-\frac{m^*_{S}}{\hbar^2}\right)\;\frac{g^2(\rho)}{\rho}\int\;d^d\vec{q}\int_{\substack{|\vec{k_1}|<k_F\\|\vec{k_1}+\vec{q}|>k_F}}d^d\vec{k_1}\int_{\substack{|\vec{k_2}|<k_F\\|\vec{k_2}-\vec{q}|>k_F}}d^d\vec{k_2}\;\frac{1}{\vec{q}^2+\vec{q}\cdot(\vec{k_1}-\vec{k_2})},
\end{eqnarray}
where $k_F$ is the Fermi momentum that we can express in terms of the density $\rho$, $k_F(\rho)=\left(\frac{3\pi^2}{2}\rho\right)^{\frac{1}{3}}$, and $m^*_S$ is the isoscalar effective mass for which we take the mean--field value (as in Ref. \cite{mog2012}). 
In the simple $t_0-t_3$ model that we consider here $m^*_S=m$ because the velocity-dependent terms of the Skyrme force are missing.

In our case the regulator $\epsilon$ can be written in terms of the dimension $d$ as $\epsilon=3-d$. When $\epsilon\rightarrow0$, $d$ returns to the integer value 3.  
\end{widetext}
By making some manipulations and by  using the Schwinger's proper time representation of Feynman integrals the following compact expression may be derived, 

\begin{eqnarray}
\frac{\Delta E(\rho)}{A}= \tilde{C}_{d}(\mu)\;m^*_S\; k_F^{3d-5} \;g^2(\rho)\int_{C_I}\;d^d\vec{q}\;\;J_d(q). 
\end{eqnarray}
The coefficient $\tilde{C}_d$ has the following expression: 
\begin{equation}
\tilde{C}_{d}(\mu)= -\frac{1}{\hbar^2} \frac{\Omega^{d-3}}{\mu^{3(d-3)}}\frac{3\pi^2}{(2\pi)^{3d}}. 
\end{equation}
We observe that the coefficient $C_d$ is negative. 
The domain of integration in Eq. (2) is (with a rescaling in momenta):
\begin{equation}
C_I=\left[|\vec{k_1}|,|\vec{k_2}|<1;\;|\vec{k_1}+\vec{q}|,|\vec{k_2}-\vec{q}|>1\right].
\end{equation}
The integrand in Eq. (2) can be written as
\begin{equation}
J_d(q)=\int_0^{\infty}\;d\alpha\;e^{-\alpha q^2}\;\left[I_d(\alpha,q)\right]^2, 
\end{equation}
where
\begin{equation}
I_d(\alpha,q)=\int_{\substack{|\vec{k}|<1\\|\vec{k}+\vec{q}|>1}}d^d\vec{k}\;e^{-\alpha\vec{q}\cdot\vec{k}}.
\end{equation}

In dimension $d$, all massless integrals are regularized to zero \cite{leibb,mitra,gh}. 
For instance, it holds: 
\begin{eqnarray}
I=\int\;d^d\vec{q}\int_{|k_1|,|k_2|<1} d^d\vec{k_1}\;d^d\vec{k_2}\;\frac{1}{q^2}=0.\quad 
\end{eqnarray}
 
By rewriting $I$ as the sum of two integrals $I_1$ and $I_2$ (by splitting the integration in two regions),  

\begin{eqnarray}
\nonumber
I_1&=&\int_{|q|>2}\;d^d\vec{q}\int_{|k_1|,|k_2|<1} d^d\vec{k_1}\;d^d\vec{k_2}\;\frac{1}{q^2},
\\
I_2&=&\int_{|q|<2}\;d^d\vec{q}\int_{|k_1|,|k_2|<1} d^d\vec{k_1}\;d^d\vec{k_2}\;\frac{1}{q^2}, 
\end{eqnarray}
the second--order correction reads 

\begin{widetext}
\begin{eqnarray}
\nonumber 
\frac{\Delta E}{A}(\rho)&=&C_{d}(\mu,k_F(\rho))\;g^2(\rho)\left[\left(\int_{|q|>2}\;d^d\vec{q}\;J_d(q)\;-I_1\right)
+\left(\int_{|q|<2}\;d^d\vec{q}\;J_d(q)\;-I_2\right)\right] \\
&=& C_{d}(\mu,k_F(\rho))\;g^2(\rho)\left[A+B\right]. 
\end{eqnarray}
\end{widetext}
The quantity $B$ is finite for $|q|<2$ when $d \rightarrow 3$ 
and one can show that its value is equal in this case to
\begin{equation}
64\pi^3\left(\frac{59}{315}-\frac{46}{105}\ln 2\right). 
\end{equation} 
The quantity $A$ may be written as a function of $d$,\newpage $A=\frac{\pi^d}{\left[\Gamma\left(1+\frac{d}{2}\right)\right]^2}T(d)$. By using hypergeometric functions \cite{Larry}, 
after straightforward manipulations one can write 
\begin{widetext}
\begin{eqnarray}
T(d)=\frac{2^{d-1}\pi}{1-\frac{d}{2}}\left[{}_{4}\text{F}_{3}\left(\frac{1+d}{2},\frac{1}{2},1,\frac{2-d}{2};1+\frac{d}{2},1+d,\frac{4-d}{2};1\right)-1\right], \;\;\; 
\lim_{d\rightarrow 3}T(d)=16\pi\left(-\frac{23}{35}+\frac{36}{35}\ln 2\right). 
\end{eqnarray} 
\end{widetext}

$T(d)$ converges for $0\le d<4$ (with a pole at $d= 4$) as shown in Fig. 1. For $d=3$ the divergence 
has been removed by the regularization procedure (the value of $T(3)$ is positive and finite).  

\begin{figure}[htbp]
\includegraphics[scale=0.9]{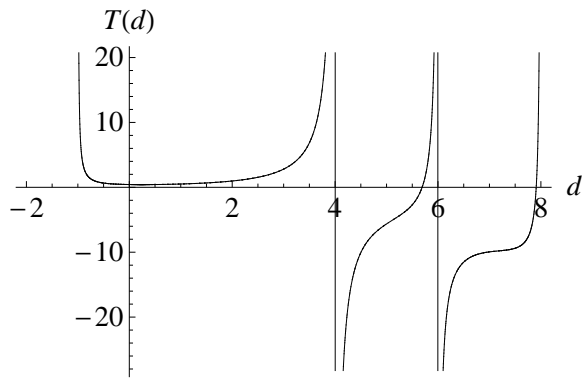} 
\caption{ $T(d)$ as a function of the space-dimension $d$.}
\end{figure}
By taking $d=3$, 
the mean--field + second--order EoS finally reads

\begin{widetext}
\begin{eqnarray}\label{eqn10}
\frac{E}{A}(\rho)=\frac{3\hbar^2}{10\;m}\left(\frac{3\pi^2}{2}\rho\right)^{\frac{2}{3}}+\frac{3}{8}\rho\;g(\rho)+
48\;\pi^3 \tilde{C} \; k_F^4(\rho) \; m^*_S\;g^2(\rho)\left(\frac{-11+2\ln 2}{105}\right), 
\end{eqnarray}
\end{widetext}
where $\tilde{C}=\tilde{C}_{d=3}$, 
\begin{equation}
\tilde{C} = \tilde{C}_{d=3} = -\frac{1}{\hbar^2} \frac{3}{(2\pi)^{7}}. 
\end{equation}
The above result for the regularized second--order correction coincides with the previous findings of Refs. \cite{hammer,kaiser}. 
After regularization, the dependence on $\mu$ of the coefficient $\tilde{C}$ has disappeared (due to the minimal subtraction that has been applied). 
We observe that, as expected, the power--law divergent terms have been removed and do not appear in the final expression. 

\subsection{ Regularized EoS for symmetric, asymmetric, and pure neutron matter: full Skyrme interaction}

By following the same procedure (by using the properties of massless integrals in dimension $d$) as for the 
case $t_0-t_3$, it is possible to regularize the second-order term also for the full Skyrme case 
where the velocity--dependent terms are included. The parameters of the interaction are nine: 
$t_0$, $t_1$, $t_2$, $t_3$, $x_0$, $x_1$, $x_2$, $x_3$, and $\alpha$. 
For simplicity, we do not report here the long analytical derivation and we provide  directly the final 
regularized expressions for the three 
specific cases that we are going to 
consider in the applications, namely, symmetric, pure neutron, and asymmetric matter. 

The first + second--order EoS for symmetric matter  reads 
\begin{eqnarray}
\nonumber
\frac{E}{A}(\rho)&=&\frac{3\hbar^2}{10m}\left(\frac{3\pi^2}{2}\rho \right)^{\frac{2}{3}} + 
\frac{3}{8} t_0 \rho+ \frac{1}{16} t_3 \rho^{\alpha+1}  \\ 
&+&\frac{3}{80}\left(\frac{3\pi^2}{2}\right)^{\frac{2}{3}}
\rho^{\frac{5}{3}}\Theta_s +\frac{\Delta E^S}{A} (\rho),
\end{eqnarray}
where
\begin{equation}
\Theta_s = 3t_1 + t_2 (5+4x_2), 
\end{equation}
and the second--order correction is given by: 
\begin{widetext}
\begin{equation}
\frac{\Delta E^S}{A} (\rho)=\frac{-11+2\ln{2}}{105}\;\chi^S_1(\rho)+\frac{-167+24 \ln{2}}{945}\;\chi^S_2(\rho)+\frac{-2066+312\ln{2}}{31185}\;\chi^S_3(\rho)+\frac{-9997+1236 \ln{2}}{62370}\;\chi^S_4(\rho).
\end{equation}
\end{widetext}
The four density--dependent coefficients $\chi^S$ have the following expressions as a function of the parameters of the interaction:
\begin{eqnarray}
\nonumber
\chi^S_1(\rho)&=&48\;\pi^3\; \tilde{C}\;m^*_S\;k_F^4(\rho)\left(t_{03}^2+x_{03}^2\right),\\
\nonumber
\chi^S_2(\rho)&=&64\;\pi^3\; \tilde{C}\;m^*_S\;k_F^6(\rho)\left(t_{03}\;t_{12}+x_{03}\;x_{12}\right)\\
\nonumber
\chi^S_3(\rho)&=&64\;\pi^3\; \tilde{C}\;m^*_S\;k_F^{8}(\rho)\left(t_{12}\;x_{12}\right), \\
\chi^S_4(\rho)&=&64\;\pi^3\; \tilde{C}\;m^*_S\;k_F^{8}(\rho)\left(t_{12}^2+\;x_{12}^2\right),
\end{eqnarray}
where $\tilde{C}$ has been defined above.   
The following notation has been used: 
\begin{eqnarray}
\nonumber
t_{03} &=& t_0 + \frac{t_3}{6} \rho^{\alpha}, \\
\nonumber
x_{03} &=& t_0 x_0 + \frac{t_3 x_3}{6} \rho^{\alpha}, \\
\nonumber
t_{12} &=& t_1 + t_2, \;\;\;\; x_{12} = t_1 x_1 + t_2 x_2. 
\end{eqnarray}
One can notice that each of the four terms in Eq. (15) has a different  $k_F$ dependence, the dependence of the first term being that of the $t_0-t_3$ model. It is easy to show that the $t_0-t_3$ result for the second--order 
correction is recovered by considering the first term where the parameters $x_0$ and $x_3$ are taken equal to zero. 

The beyond mean--field EoS evaluated at second order for pure neutron matter reads:
\begin{eqnarray}
\nonumber
\frac{E}{A}(\rho)&=& 
\frac{3\hbar^2}{10m}\left(3\pi^2\rho \right)^{\frac{2}{3}}+\frac{1}{4} t_0\left(1-x_0\right)\rho 
+ \frac{1}{24}\left(1-x_3 \right) t_3 \rho^{\alpha+1} \\ &+&\frac{3}{40}\left(3\pi^2\right)^{\frac{2}{3}}
\rho^{\frac{5}{3}}\left(\Theta_s-\Theta_v\right) +\frac{\Delta E^N}{A} (\rho), 
\end{eqnarray}
where 
\begin{equation}
\Theta_v = t_1 (2+x_1)+t_2(2+x_2). 
\end{equation}
This time the second--order correction is given by: 
\begin{widetext}
\begin{eqnarray}
\nonumber
\frac{\Delta E^N}{A} (\rho)&=&\frac{-11+2\ln{2}}{105}\;\chi^N_1(\rho)+\frac{-167+24 \ln{2}}{2835}\;\chi^N_2(\rho)+\frac{167-24 \ln{2}}{5670}\;
\chi^N_3(\rho) \\ 
&+&\frac{461-24\ln{2}}{31185}\;\chi^N_4(\rho)+\frac{-4021+516\ln{2}}{124740}\;\chi^N_5(\rho)
\end{eqnarray}
\end{widetext}
and the five density--dependent coefficients $\chi^N$ are:
\begin{eqnarray}
\nonumber
\chi^N_1(\rho)&=&8\;\pi^3\;\tilde{C}\;m^*_N\;k_N^4(\rho)\left(t_{03}-x_{03}\right)^2,\\
\nonumber
\chi^N_2(\rho)&=&32\;\pi^3\;\tilde{C}\;m^*_N\;k_N^6(\rho)\left(t_{03}\;t_{12}+x_{03}\;x_{12}\right),\\
\nonumber
\chi^N_3(\rho)&=&64\;\pi^3\; \tilde{C}\;m^*_N\;k_N^6(\rho)\left(t_{03}\;x_{12}+x_{03}\;t_{12}\right),\\
\nonumber
\chi^N_4(\rho)&=&64\;\pi^3\; \tilde{C}\;m^*_N\;k_N^{8}(\rho)\left(t_{12}\;x_{12}\right),\\
\chi^N_5(\rho)&=&64\;\pi^3\; \tilde{C}\;m^*_N\;k_N^{8}(\rho)\left(t_{12}^2+\;x_{12}^2\right).
\end{eqnarray}
In Eq. (21) $k_N$ is neutron Fermi momentum, 
$k_N(\rho)=k_F(\rho)(1+\delta)^{1/3}$, where $\delta$ is the asymmetry parameter that we can express 
in terms of the neutron and proton densities $\rho_N$ and $\rho_P$, 
\begin{equation}
\delta=\frac{\rho_N-\rho_P}{\rho_N+\rho_P}; 
\end{equation}
$\delta=1$ in this case of pure neutron matter;  
$m_N^*$ is the neutron 
effective mass (that we take equal to its mean--field value like in Ref. \cite{mog2012}). 

By using the asymmetry parameter $\delta$, the EoS for asymmetric matter reads: 
\begin{widetext}
\begin{eqnarray}
\nonumber
\frac{E}{A}(\delta,\rho) &=& 
\frac{3\hbar^2}{10m}\left(\frac{3\pi^2}{2}\rho \right)^{\frac{2}{3}} G_{5/3} + 
\frac{1}{8} t_0 \rho [2(2+x_0)-(1+2x_0)G_2] 
+ \frac{1}{48} t_3 \rho^{\alpha+1} [2(2+x_3)-(1+2x_3)G_2] \\ &+& 
\frac{3}{40}\left(\frac{3\pi^2}{2}\right)^{\frac{2}{3}}
\rho^{\frac{5}{3}}\left[ \Theta_v G_{5/3} + \frac{1}{2} (\Theta_s -2 \Theta_v) 
G_{8/3} \right] + \frac{\Delta E^{AS}(\delta, \rho)}{A},  
\end{eqnarray}
\end{widetext}
where  
\begin{equation}
G_{\beta} = \frac{1}{2} [(1+\delta)^{\beta}+(1-\delta)^{\beta}].
\end{equation}
The expression for $\Delta E^{AS}(\rho,\delta)/A$ is reported in Appendix A. 

\section{Results. Fit of the parameters}

\subsection{$t_0-t_3$ model}

As in Ref. \cite{mog2010} we use the SkP \cite{doba} mean--field EoS as a reference EoS for the $t_0-t_3$ model and we consider 
only symmetric nuclear matter.
We plot in Fig. 2 the reference mean-field EoS for symmetric nuclear matter together with the first + second--order 
EoS calculated with the same parameters (a). In panel (b) only the second-order corrective term is plotted. 
One can notice that the regularized second-order correction is always positive and has the opposite curvature with respect to the first--order EoS. 
This occurs because in the simple 
$t_0-t_3$ model the second--order term is just proportional to the square of the coupling constant $g$ and the multiplying factor is positive because the coefficient $C_3$ is negative (see Eq. (12)).  
\begin{figure}[htbp]
\includegraphics[scale=0.35]{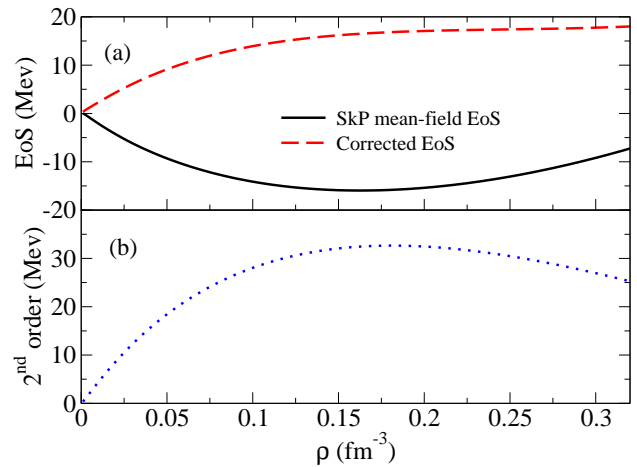} 
\caption{ Colors online. (a) SkP mean--field EoS (full line) and mean--field + second--order EoS calculated with the SkP parameters (dashed line) for the $t_0-t_3$ model in symmetric matter. (b) Second-order correction.}
\end{figure}
The second--order correction has its maximum value (due to the change of curvature, the second-order term has a 
maximum) around the equilibrium density for matter.  
The corrected EoS is strongly 
modified with respect to the mean--field EoS calculated with the same parameters and, in particular, 
the curvature is changed because the second--order correction is dominant. 
Due to the difference of curvature it turns out that 
the adjustment of the corrected EoS to a reasonable reference EoS (we have chosen
 the SkP mean--field EoS as a reference) is not feasible: The minimization of a $\chi^2$ does not lead to any result and this means that 
the parameters cannot be adjusted in this case. We have checked that a fit would be possible only in some windows of density   
at large values, well beyond the saturation density.

\subsection{Full Skyrme interaction}

In the case of the full Skyrme interaction the reference mean--field EoS (on which the fit is performed) has been evaluated with the parametrization SLy5 \cite{chaba} as in Ref. \cite{mog2012}.
 All the nine parameters of the interaction are kept free in the fitting procedure. 
We have first performed separate fits for the single cases of symmetric, asymmetric and pure neutron matter. As an illustration, 
we present in what follows the results corresponding to symmetric and pure neutron matter.
Then, we have performed some simultaneous fits that will be presented in the last part of this section.  
The following $\chi^2$ is minimized: 
\begin{equation}
\chi^2 = \frac{1}{N-1} \sum_{i=1}^N \frac{(E_i-E_{i,ref})^2}{\Delta E^2_i}.  
\end{equation}
The number $N$ of fitted points is 15  and the errors $\Delta E_i$ are chosen equal to 1\% of the reference SLy5 mean-field energies 
$E_{i,ref}$. The points are in the range of densities between 0.02 to 0.30 fm$^{-3}$. In the case of the 
simultaneous fit of symmetric and pure neutron matter the points are in the 
range of densities between 0.1 and 0.3  fm$^{-3}$.

\vspace{1cm}

{\it Symmetric matter.} 

The corrected second--order EoS is plotted in Fig. 3 together with the reference mean--field EoS. We can observe that the correction is extremely 
large 
also in this case 
but the curvature is not modified with respect to the mean--field EoS. This is due to the  
analytical expression of the 
second-order correction (in this case there is no a simple dependence on the square of a coupling constant as in the $t_0-t_3$ model). The  number of parameters is large enough to perform successfully the fit.

\begin{figure}[htbp]
\includegraphics[scale=0.35]{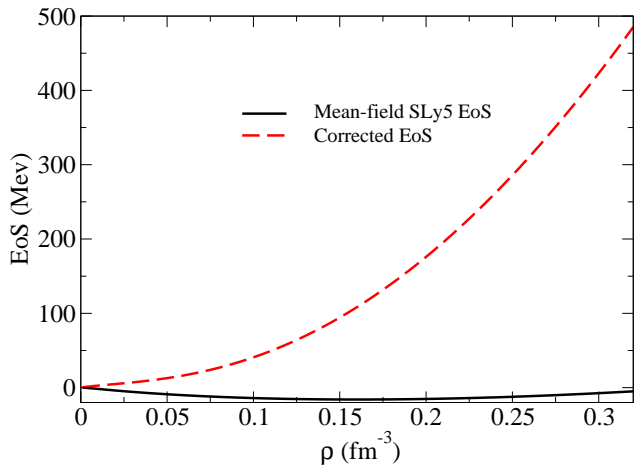} 
\caption{ (Color online) SLy5 mean--field EoS (full line) and mean--field + second--order EoS calculated with the SLy5 parameters (dashed line) for 
symmetric matter. }
\end{figure}

\begin{figure}[htbp]
\includegraphics[scale=0.35]{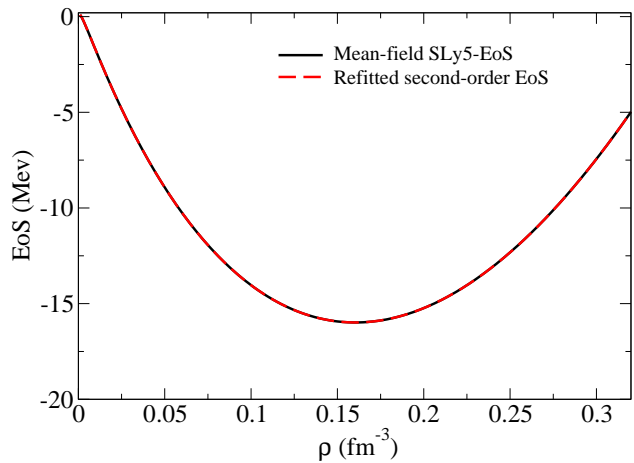} 
\caption{ (Color online)  SLy5 mean--field EoS (full line) and refitted second--order EoS (dashed line) for symmetric matter.}
\end{figure}

\begin{figure}[htbp]
\includegraphics[scale=0.35]{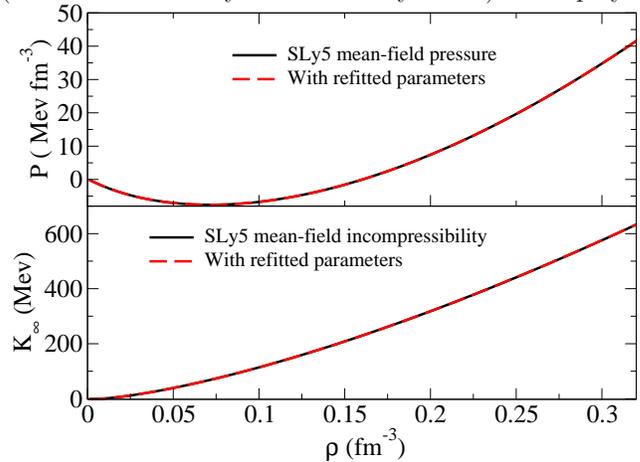} 
\caption{ (Color online) (a) Mean--field SLy5 pressure (full line) and second--order pressure obtained with the refitted parameters of Table I (dashed line). (b) Same as in (a) but for the incompressibility. }
\end{figure}

The results of the fit are shown in Fig. 4 and the corresponding second--order pressure and incompressibility (that are not directly constrained by the fit) are displayed in Fig. 5. In particular, the incompressibility value at saturation density is equal to 229.5 MeV. 
The parameters and the corresponding $\chi^2$ value are reported in Table I (third line). 
The quality of the fit is extremely good as one can deduce by the figures and by the small value of the corresponding $\chi^2$. 
We have then performed a final check with the refitted parameters: 
we have computed the symmetry energy $a_{\delta}$ 
by using the second derivative of the second--order EoS for asymmetric matter. 
 We have been obtained $a_I= 1643.3 $ MeV that is  very far 
from 
 the range of acceptable values. For this reason, 
we have added in the fitting precedure an additional constraint on the value of the symmetry energy (by using as a reference the mean--field SLy5 value). We do not present here the corresponding plots because they are very similar to Figs. 4 and 5 (the incompressibility value at saturation density is now equal to 228.5 MeV). We report  
in the fourth line of Table I the new parameters. We observe that the value of $\chi^2$ is larger than in the previous fit 
indicating that the inclusion of the new constraint slightly deteriorates the quality of the fit that is still anyway extremely good. 
The symmetry energy is now equal to 32.03 MeV (equal to the mean-field SLy5 value). 

\begin{widetext}

\begin{table}[!h]
\centering
\caption{Parameter sets obtained in the fit of the EoS of symmetric matter compared with the original set SLy5. In the last column the $\chi^2$ values are shown. In the last line the parameters correspond to the fit where an additional constraint on the symmetry energy value is added.}\vspace{0.5cm}
\begin{tabular}{ c c c c c c c c c c c c }
    \hline
     &    $\quad t_0$ & $t_1$ & $t_2$ & $t_3$ & $x_0$ & $x_1$ & $x_2$ & $x_3$ & $\alpha$ &$\chi^2$&\\
     &    (MeV fm$^3$) & (MeV fm$^5$) & MeV fm$^5$) & (MeV fm$^{3+3\alpha}$) & & & & & &  \\
\hline
SLy5 &-2484.88    &    483.13   &   -549.40   & 13763.0&0.778     &    -0.328    &  -1.0 &1.267   &   0.16667  &--&\\   
\hline
New&-2510.87& 20239.43& -897.06& -1176280.24&  0.065 & -1.272 &-21.775& -0.656&  0.663 & 3.5$\times 10^{-7}$ &\\
\hline
New$_{a_I}$ & -3401.65 &  28666.59 & -970.62 & -1938032.85  & 0.330 & -1.563 & -24.078 & -0.819 & 0.666 &  4.6$\times 10^{-3}$ & \\
   \hline
  \end{tabular}
  \end{table}
  
\end{widetext}

\vspace{1cm}

{\it Neutron matter.}

The mean--field SLy5 EoS and the second-order EoS (calculated with the SLy5 parameters) are plotted in Fig. 6 for pure neutron matter. 
The $\chi^2$ minimizaton provides the result that is displayed in Fig. 7 and the corresponding parameters are given in Table II.
Also in this case the quality of the fit is extremely good (the value of the corresponding $\chi^2$ is very small).

\begin{figure}[htbp]
\includegraphics[scale=0.35]{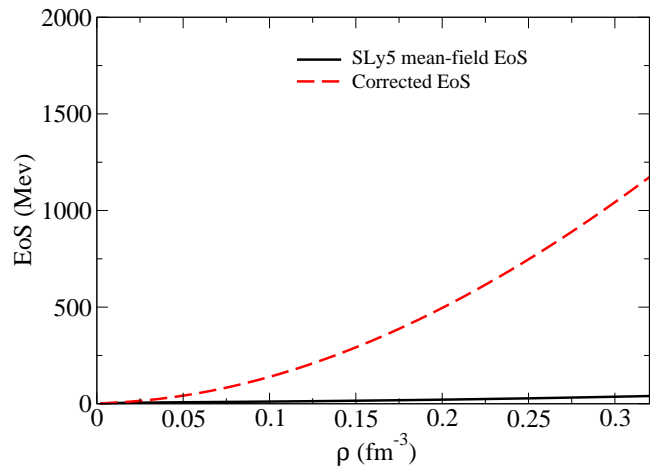} 
\caption{ (Color online) SLy5 mean--field EoS (full line) and mean--field + second--order EoS calculated with the SLy5 parameters (dashed line) for neutron matter. }
\end{figure}

\begin{figure}[htbp]
\includegraphics[scale=0.35]{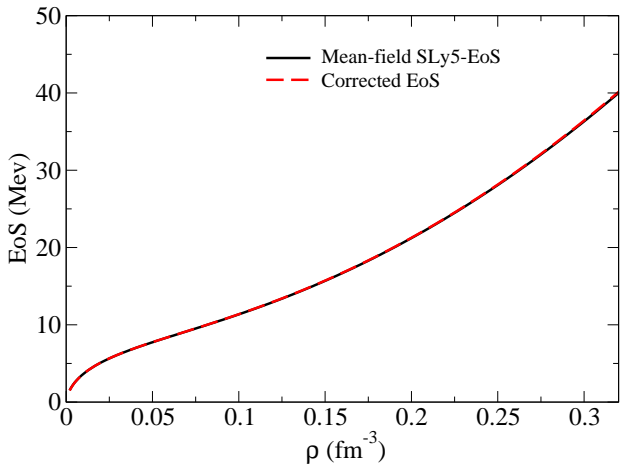} 
\caption{ (Color online) SLy5 mean--field EoS (full line) and refitted second--order EoS (dashed line) for neutron matter (parameters of Table II). }
\end{figure}

\begin{widetext}

\begin{table}[!h]
\centering
\caption{Parameter sets obtained in the fit of the EoS of pure neutron matter compared with the original set SLy5. In the last column the $\chi^2$ value is shown. }\vspace{0.5cm}
\begin{tabular}{ c c c c c c c c c c c c }
    \hline
     &    $\quad t_0$ & $t_1$ & $t_2$ & $t_3$ & $x_0$ & $x_1$ & $x_2$ & $x_3$ & $\alpha$ &$\chi^2$&\\
     &    (MeV fm$^3$) & (MeV fm$^5$) & MeV fm$^5$) & (MeV fm$^{3+3\alpha}$) & & & & & & & \\ 
\hline
SLy5 &-2484.88    &    483.13   &   -549.40   & 13763.0&0.778     &    -0.328    &  -1.0 &1.267   &   0.16667  &--&\\   
\hline
New&-3287.287& 2038.711& -459.159&  109814.050&  0.706& -1.645& -3.861& 3.127&  0.656& 1.15$\times 10^{-4}$& \\
   \hline
  \end{tabular}
  \end{table}
  
 \end{widetext}

{\it Global fits}

We have first tried as in our previous work \cite{mog2012} to perform a global fit 
by including simultaneously 
 symmetric,  pure neutron matter, and also a case of asymmetric matter. 
It turns out that the minimization of the $\chi^2$ in this case does not provide any result. 
We have thus performed global fits by including only two equations of state. 

In Fig. 8 we present the global fit performed by considering together symmetric and pure neutron matter. The parameters are given in Table III and the resulting pressure and 
incompressibility are plotted in Fig. 9. The incompressibility value at saturation density is equal this time to 252.18 MeV. 
The $\chi^2$ value is larger than in the previous cases indicating that the quality of the fit has been deteriorated (but it still remains reasonably good). This can be seen also by looking at the plotted curves.  

\begin{widetext}

\begin{table}[!h]
\centering
\caption{Parameter sets obtained in the fit of the EoS of symmetric and pure neutron matter compared with the original set SLy5. In the last column the $\chi^2$ value is shown. }\vspace{0.5cm}
\begin{tabular}{ c c c c c c c c c c c c }
    \hline
     &    $\quad t_0$ & $t_1$ & $t_2$ & $t_3$ & $x_0$ & $x_1$ & $x_2$ & $x_3$ & $\alpha$ &$\chi^2$&\\
     &    (MeV fm$^3$) & (MeV fm$^5$) & MeV fm$^5$) & (MeV fm$^{3+3\alpha}$) & & & & & & & \\ 
\hline
SLy5 &-2484.88    &    483.13   &   -549.40   & 13763.0 & 0.778     &    -0.328    &  -1.0 &1.267   &   0.16667  &--&\\   
\hline
New& -460.73 &  10403.66 &  -8485.73  &  -141558.6 &   1.460 &  -0.681 & -0.641 & -0.779 &  0.650 &   0.202 & \\
   \hline
  \end{tabular}
  \end{table}
  
 \end{widetext}

\begin{figure}[htbp]
\includegraphics[scale=0.35]{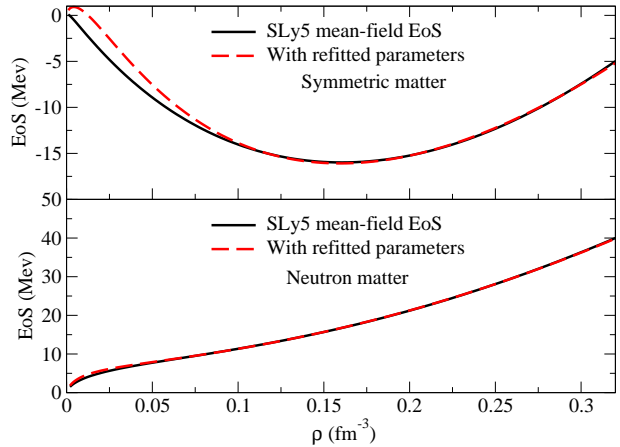} 
\caption{ (Color online) (a) SLy5 mean--field (full line) and refitted (dashed line) EoS for symmetric matter. (b) Same as in (a) but for neutron matter. The results are obtained by fitting simultaneously 
symmetric and neutron matter (parameters of Table III).}
\end{figure}

\begin{figure}[htbp]
\includegraphics[scale=0.35]{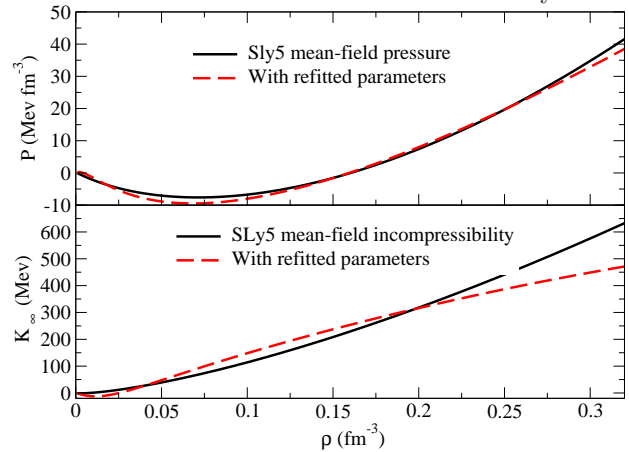} 
\caption{ (Color online) (a) Mean--field SLy5 pressure (full line) and second--order pressure obtained with the refitted parameters (dashed line). (b) Same as in (a) but for the incompressibility. The results are obtained by fitting simultaneously 
symmetric and neutron matter (parameters of Table III).}
\end{figure}

 We have computed also in this case the  symmetry energy (by using the second--order EoS  for asymmetric matter) with the  parameters of Table III and obtained the value of  -7.5 MeV that is not acceptable. 
We have partially succedeed this time in adding also a constraint on the value of the symmetry energy: 
what we have found with this last fit is a value of 32.7 MeV for the symmetry energy and a value of 310.74 MeV for the incompressibility at the saturation point. However, the associated $\chi^2$ is equal to 4.49, that indicates that the fit is much less good than in the previous cases. 
The fitted curves (we do not show the curves and the parameters for this case) are of much lower quality than for the other fits. 
We conclude that the global fit that includes symmetric and pure neutron matter does not provide satisfactory results:
either it does not lead to a reasonable value for the symmetry energy or it is of quite low quality (when the constraint on the symmetry energy is explicitly introduced).

We have performed a new global fit by disregarding the case of pure 
neutron matter and by considering together symmetric and asymmetric matter (with $\delta=0.5$). 
The results are shown in Fig. 10 and the parameters are listed in Table IV. 

\begin{figure}[htbp]
\includegraphics[scale=0.35]{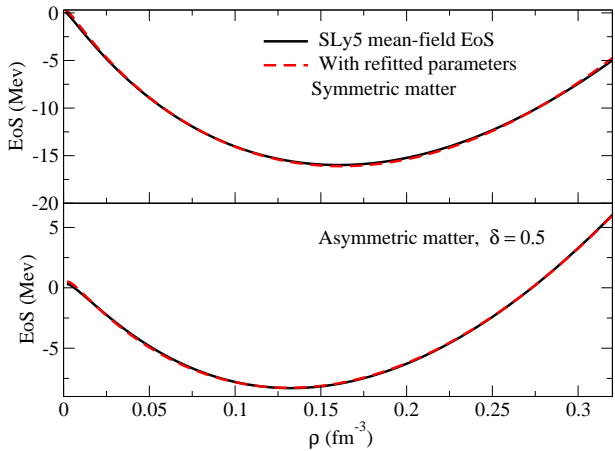} 
\caption{ (Color online)  (a) SLy5 mean--field (full line) and refitted (dashed line) EoS for symmetric matter. (b) Same as in (a) but for neutron matter. The results are obtained by fitting simultaneously 
symmetric and asymmetric matter with $\delta=0.5$ (parameters of Table IV). }
\end{figure}

\begin{figure}[htbp]
\includegraphics[scale=0.35]{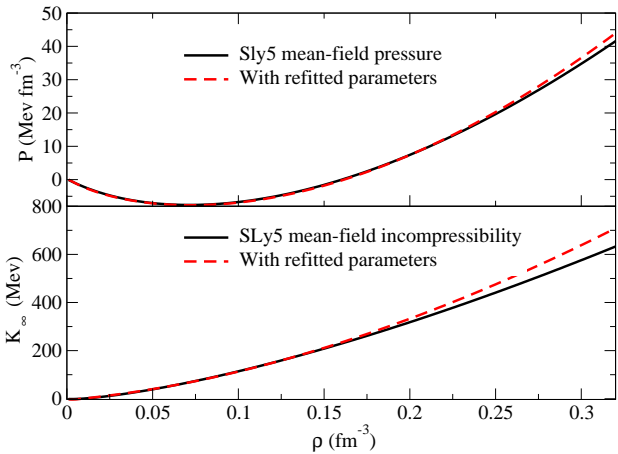} 
\caption{ (Color online) (a) Mean--field SLy5 pressure (full line) and second--order pressure obtained with the refitted parameters (dashed line). (b) Same as in (a) but for the incompressibility. The results are obtained by fitting simultaneously 
symmetric and asymmetric matter with $\delta=0.5$ (parameters of Table IV). }
\end{figure}

\begin{widetext}

\begin{table}[!h]
\centering
\caption{Parameter sets obtained in the fit of the EoS of symmetric and asymmetric matter compared with the original set SLy5. 
In the last column the $\chi^2$ value is shown. In the last line the parameters correspond to the fit where 
an additional constraint on the symmetry energy value has been added. }\vspace{0.5cm}
\begin{tabular}{ c c c c c c c c c c c c }
    \hline
     &    $\quad t_0$ & $t_1$ & $t_2$ & $t_3$ & $x_0$ & $x_1$ & $x_2$ & $x_3$ & $\alpha$ &$\chi^2$&\\
     &    (MeV fm$^3$) & (MeV fm$^5$) & MeV fm$^5$) & (MeV fm$^{3+3\alpha}$) & & & & & & & \\ 
\hline
SLy5 &-2484.88    &    483.13   &   -549.40   & 13763.0 & 0.778     &    -0.328    &  -1.0 &1.267   &   0.16667  &--&\\   
\hline
New& -1653.09 &   10346.70 &  -698.66 &  -784064.9 &  0.716 &    -1.715 &    -14.9 &   -0.697 &   0.667 &     1.337  &\\
   \hline
New$_{a_I}$ &  -1656.03 & 10526.54 &  -727.53 &  -795583.4 &   0.729 &  -1.783 &  -14.7 &  -0.725 &  0.667 &   1.341 & \\
\hline
  \end{tabular}
  \end{table}
  
 \end{widetext}

The pressure and incompressibility values are plotted in Fig 11 (the incompressibility at saturation is equal to 233.8 MeV). 
The symmetry energy is equal to 24.9 MeV.  
If we include in the fitting procedure an additional constraint on the symmetry energy value we obtain curves that are very similar to those already shown in Fig. 10, a value for the incompressibility at the saturation point equal to 233.9 MeV and a value for the symmetry energy equal to 32.00 MeV. The corresponding new parameters are reported in the last line of Table IV and are not very different with respect to the previous set that is shown in the line above.

\section{Conclusions}

In the present work we have regularized the second--order divergent term of the EoS of nuclear matter evaluated with a zero--range Skyrme interaction. The dimensional regularization technique has been used. Applications of this technique to symmetric matter within a $t_0-t_3$ model and to different types of matter with the full Skyrme interaction have been presented. In particular, for this last case, it has been shown that a fit of the regularized EoS to a reference EoS is possible. 
Symmetric, pure neutron and asymmetric matter can be reproduced separately. It is also shown that simultaneous fits of symmetric and neutron matter or of 
symmetric and asymmetric (with $\delta=0.5$) matter can be done. 
It has been verified that particular attention must be payed to generate sets of parameters that lead to reasonable values of the symmetry energy at the second order. For this, an additional constraint on the value of the symmetry energy can be added and satisfactory results are found in this way in the cases where symmetric matter or symmetric and asymmetric matter (simultaneously) are fitted.  

These encouraging results 
open new prospectives for future applications of 
dimensional regularized  
Skyrme-type effective interactions adjusted at a beyond mean-field level. These interactions would be well adapted to be used in the framework of beyond mean--field approaches for finite nuclei.


\noindent{\bf Acknowledgments}
We thank Gianluca Col\`o and Bira van Kolck for valuable discussions. 
One of us (K.M.) thanks Haitham Zaraket for fruitful discussions.

\newpage

\begin{widetext}
\section{APPENDIX A}

Let us define the parameters $a$, $b$, and $c$: 
\begin{eqnarray}
\nonumber
a&=&\frac{k_P}{k_N}=\left(\frac{1-\delta}{1+\delta}\right)^{1/3}, \\
b&=&\frac{m_S^*}{m_N^*}, \\
\nonumber
c&=&\frac{m_S^*}{m_P^*}, 
\end{eqnarray}
where $k_P$ and $m_P^*$ are the proton Fermi momentum and the proton effective mass, respectively. 
For asymmetric matter, the beyond-mean-field equation of state evaluated at the second-order is given by:
\begin{eqnarray*}
\frac{\Delta E^{AS}}{A}(\delta,\rho) = 
\sum_{i=1}^{6}\chi^{AS}_i(\rho,\delta)\;I_i(\rho,\delta)
\end{eqnarray*}
where the six density-dependent coefficients $\chi^{AS}(\rho,\delta)$ and the factors $I_i(\rho,\delta)$ are given by:
\begin{minipage}[l]{0.55\textwidth}
\begin{eqnarray*}
\chi^{AS}_1(\rho,\delta)&=&8\;\pi^3\;\tilde{C}\;m^*_S\;k_N^4(\rho,\delta)\left(t_{03}^2+x_{03}^2\right),\\
\nonumber
\chi^{AS}_2(\rho,\delta)&=&-16\;\pi^3\;\tilde{C}\;m^*_S\;k_N^4(\rho,\delta)\left(t_{03}\;x_{03}\right),\\
\nonumber
\chi^{AS}_3(\rho,\delta)&=&32\;\pi^3\;\tilde{C}\;m^*_S\;k_N^6(\rho,\delta)\left(t_{03}\;t_{12}+x_{03}\;x_{12}\right),\\
\nonumber
\chi^{AS}_4(\rho,\delta)&=&64\;\pi^3\;\tilde{C}\;m^*_S\;k_N^6(\rho,\delta)\left(t_{03}\;x_{12}+x_{03}\;t_{12}\right),\\
\nonumber
\chi^{AS}_5(\rho,\delta)&=&64\;\pi^3\;\tilde{C}\;m^*_S\;k_N^{8}(\rho,\delta)\left(t_{12}\;x_{12}\right),\\
\chi^{AS}_6(\rho,\delta)&=&64\;\pi^3\;\tilde{C}\;m^*_S\;k_N^{8}(\rho,\delta)\left(t_{12}^2+\;x_{12}^2\right),
\end{eqnarray*}
\end{minipage}%
\begin{minipage}[l]{0.5\textwidth}
\begin{eqnarray*}
I_1&=&\frac{-11+2\ln{2}}{105}\left(\frac{1}{b}+\frac{a^7}{c}\right)+4S_{1},\\
I_2&=&\frac{-11+2\ln{2}}{105}\left(\frac{1}{b}+\frac{a^7}{c}\right)-2S_{1},\\
I_3&=&\frac{-167+24 \ln{2}}{2835}\left(\frac{1}{b}+\frac{a^9}{c}\right)+8S_{2},\\
I_4&=&\frac{167-24 \ln{2}}{5670}\left(\frac{1}{b}+\frac{a^9}{c}\right)+2S_{2},\\
I_5&=&\frac{461-24\ln{2}}{31185}\left(\frac{1}{b}+\frac{a^{11}}{c}\right)+8S_{3},\\
I_6&=&\frac{-4021+516\ln{2}}{124740}\left(\frac{1}{b}+\frac{a^{11}}{c}\right)+8S_{3}.
\end{eqnarray*}
\end{minipage}%

The dependence on $\delta$ of the quantities $I_i$ is contained on the parameters $a$, $b$, and $c$. 
By using the dimensional regularization technique with the minimal subtraction method,  $S_1,\;S_2$ and $S_3$ result:
\begin{eqnarray*}
S_{1}&=&\frac{1}{(bc)^3}\left[\frac{1}{15}\int_{0}^{a}u\;F_1^{abc}(u)\;du+\frac{1}{15}\int_{a}^{1}u\;F_3^{abc}(u)\;du+T_1(a,b,c)\right],\\
S_{2}&=&\frac{1}{(bc)^3}\left[\frac{1}{15}\int_{0}^{a}u^3\;F_1^{abc}(u)\;du+\frac{1}{15}\int_{a}^{1}u^3\;F_3^{abc}(u)\;du+T_2(a,b,c)\right],\\
S_{3}&=&\frac{1}{(bc)^3}\left[\frac{1}{15}\int_{0}^{a}u^5\;F_1^{abc}(u)\;du+\frac{1}{15}\int_{a}^{1}u^5\;F_3^{abc}(u)\;du+T_3(a,b,c)\right].
\end{eqnarray*}
The integrals are calculated numerically. The expressions for the functions $F_1^{abc}$ and $F_3^{abc}$ are provided in Ref. \cite{mog2012} and the expressions for  $T_1(a,b,c),\;T_2(a,b,c)$ and $T_3(a,b,c)$ are, 
\begin{eqnarray*}
T_1(a,b,c)&=&-\frac{abc}{420}\left(16+88 b^2+b^4+88 a^2 c^2+22 a^2 b^2 c^2+a^4 c^4\right)+\frac{1}{3360}(2-b-a c)^4(-8-16 b+8 b^2+b^3-16 a c\\&-&40 a b c-4 a b^2 c+8 a^2 c^2-4 a^2 b c^2+a^3 c^3) \ln (2-b-a c)+\frac{1}{3360}(2+b-a c)^4 (8-16 b-8 b^2+b^3+16 a c-40 a b c\\
&+&4 a b^2 c-8 a^2 c^2-4 a^2 b c^2-a^3 c^3) \ln(2+b-a c)-\frac{1}{3360}(2-b+a c)^4(-8-16 b+8 b^2+b^3+16 a c+40 a b c\\
&+&4 a b^2 c+8 a^2 c^2-4 a^2 b c^2-a^3 c^3) \ln(2-b+a c)-\frac{1}{3360}(2+b+a c)^4(8-16 b-8 b^2+b^3-16 a c+40 a b c-4 a b^2 c\\
&-&8 a^2 c^2-4 a^2 b c^2+a^3 c^3) \ln(2+b+a c)
\end{eqnarray*}

\begin{eqnarray*}
T_2(a,b,c)&=&-\frac{1}{45360}a b c \left(1344+6096 b^2+12 b^4+3 b^6+6096 a^2 c^2+136 a^2 b^2 c^2+237 a^2 b^4 c^2+12 a^4 c^4+237 a^4 b^2 c^4+3 a^6 c^6\right)\\
&+&\frac{1}{120960}(2-b-a c)^4(-224-448 b+160 b^2+40 b^3+8 b^4+b^5-448 a c-1120 a b c-240 a b^2 c-40 a b^3 c\\
&-&4 a b^4 c+160 a^2 c^2-240 a^2 b c^2-96 a^2 b^2 c^2-17 a^2 b^3 c^2+40 a^3 c^3-40 a^3 b c^3-17 a^3 b^2 c^3+8 a^4 c^4-4 a^4 b c^4\\
&+&a^5 c^5)\ln(2-b-a c)+\frac{1}{120960}(2+b-a c)^4(224-448 b-160 b^2+40 b^3-8 b^4+b^5+448 a c-1120 a b c\\
&+&240 a b^2 c-40 a b^3 c+4 a b^4 c-160 a^2 c^2-240 a^2 b c^2+96 a^2 b^2 c^2-17 a^2 b^3 c^2-40 a^3 c^3-40 a^3 b c^3+17 a^3 b^2 c^3\\
&-&8 a^4 c^4-4 a^4 b c^4-a^5 c^5)\ln(2+b-a c)-\frac{1}{120960}(2-b+a c)^4(-224-448 b+160 b^2+40 b^3+8 b^4+b^5\\
&+&448 a c+1120 a b c+240 a b^2 c+40 a b^3 c+4 a b^4 c+160 a^2 c^2-240 a^2 b c^2-96 a^2 b^2 c^2-17 a^2 b^3 c^2-40 a^3 c^3\\
&+&40 a^3 b c^3+17 a^3 b^2 c^3+8 a^4 c^4-4 a^4 b c^4-a^5 c^5)\ln(2-b+a c)-\frac{1}{120960}(2+b+a c)^4(224-448 b-160 b^2+40 b^3\\
&-&8 b^4+b^5-448 a c+1120 a b c-240 a b^2 c+40 a b^3 c-4 a b^4 c-160 a^2 c^2-240 a^2 b c^2+96 a^2 b^2 c^2-17 a^2 b^3 c^2+40 a^3 c^3\\
&+&40 a^3 b c^3-17 a^3 b^2 c^3-8 a^4 c^4-4 a^4 b c^4+a^5 c^5)\ln(2+b+a c)
\end{eqnarray*}
\begin{eqnarray*}
T_3(a,b,c)&=&-\frac{1}{3991680}a b c(96768+393792 b^2+240 b^4+60 b^6+15 b^8+393792 a^2 c^2+2208 a^2 b^2 c^2+2180 a^2 b^4 c^2+2820 a^2 b^6 c^2\\
&+&240 a^4 c^4+2180 a^4 b^2 c^4+7770 a^4 b^4 c^4+60 a^6 c^6+2820 a^6 b^2 c^6+15 a^8 c^8)+\frac{1}{10644480}(2-b-a c)^4(-16128\\
&-&32256 b+8960 b^2+2800 b^3+800 b^4+200 b^5+40 b^6+5 b^7-32256 a c-80640 a b c-22400 a b^2 c-5600 a b^3 c\\
&-&1200 a b^4 c-200 a b^5 c-20 a b^6 c+8960 a^2 c^2-22400 a^2 b c^2-12800 a^2 b^2 c^2-4600 a^2 b^3 c^2-1160 a^2 b^4 c^2-170 a^2 b^5 c^2\\
&+&2800 a^3 c^3-5600 a^3 b c^3-4600 a^3 b^2 c^3-1840 a^3 b^3 c^3-375 a^3 b^4 c^3+800 a^4 c^4-1200 a^4 b c^4-1160 a^4 b^2 c^4-375 a^4 b^3 c^4\\
&+&200 a^5 c^5-200 a^5 b c^5-170 a^5 b^2 c^5+40 a^6 c^6-20 a^6 b c^6+5 a^7 c^7)\ln(2-b-a c)\\
&+&\frac{1}{10644480}(2+b-a c)^4(16128-32256 b-8960 b^2+2800 b^3-800 b^4+200 b^5-40 b^6+5 b^7+32256 a c-80640 a b c\\
&+&22400 a b^2 c-5600 a b^3 c+1200 a b^4 c-200 a b^5 c+20 a b^6 c-8960 a^2 c^2-22400 a^2 b c^2+12800 a^2 b^2 c^2-4600 a^2 b^3 c^2\\
&+&1160 a^2 b^4 c^2-170 a^2 b^5 c^2-2800 a^3 c^3-5600 a^3 b c^3+4600 a^3 b^2 c^3-1840 a^3 b^3 c^3+375 a^3 b^4 c^3-800 a^4 c^4-1200 a^4 b c^4\\
&+&1160 a^4 b^2 c^4-375 a^4 b^3 c^4-200 a^5 c^5-200 a^5 b c^5+170 a^5 b^2 c^5-40 a^6 c^6-20 a^6 b c^6-5 a^7 c^7)\ln(2+b-a c)\\
&-&\frac{1}{10644480}(2-b+a c)^4(-16128-32256 b+8960 b^2+2800 b^3+800 b^4+200 b^5+40 b^6+5 b^7+32256 a c\\
&+&80640 a b c+22400 a b^2 c+5600 a b^3 c+1200 a b^4 c+200 a b^5 c+20 a b^6 c+8960 a^2 c^2-22400 a^2 b c^2-12800 a^2 b^2 c^2\\
&-&4600 a^2 b^3 c^2-1160 a^2 b^4 c^2-170 a^2 b^5 c^2-2800 a^3 c^3+5600 a^3 b c^3+4600 a^3 b^2 c^3+1840 a^3 b^3 c^3+375 a^3 b^4 c^3\\
&+&800 a^4 c^4-1200 a^4 b c^4-1160 a^4 b^2 c^4-375 a^4 b^3 c^4-200 a^5 c^5+200 a^5 b c^5+170 a^5 b^2 c^5+40 a^6 c^6-20 a^6 b c^6\\
&-&5 a^7 c^7)\ln(2-b+a c)-\frac{1}{10644480}(2+b+a c)^4(16128-32256 b-8960 b^2+2800 b^3-800 b^4+200 b^5-40 b^6\\
&+&5 b^7-32256 a c+80640 a b c-22400 a b^2 c+5600 a b^3 c-1200 a b^4 c+200 a b^5 c-20 a b^6 c-8960 a^2 c^2-22400 a^2 b c^2\\
&+&12800 a^2 b^2 c^2-4600 a^2 b^3 c^2+1160 a^2 b^4 c^2-170 a^2 b^5 c^2+2800 a^3 c^3+5600 a^3 b c^3-4600 a^3 b^2 c^3+1840 a^3 b^3 c^3\\
&-&375 a^3 b^4 c^3-800 a^4 c^4-1200 a^4 b c^4+1160 a^4 b^2 c^4-375 a^4 b^3 c^4+200 a^5 c^5+200 a^5 b c^5-170 a^5 b^2 c^5-40 a^6 c^6\\
&-&20 a^6 b c^6+5 a^7 c^7)\ln(2+b+a c)
\end{eqnarray*}
\end{widetext}

\end{document}